\documentclass{article}

\usepackage[preprint,nonatbib]{neurips_2019}

\usepackage[utf8]{inputenc} 
\usepackage[T1]{fontenc}    
\usepackage{hyperref}       
\usepackage{url}            
\usepackage{booktabs}       
\usepackage{amsfonts}       
\usepackage{nicefrac}       
\usepackage{microtype}      
\usepackage{amsmath}
\usepackage{graphicx}
\graphicspath{ {images/} }

\title{MelGAN-VC: Voice Conversion and Audio Style Transfer on arbitrarily long samples using Spectrograms}

\author{
  Marco Pasini\\
  Alma Mater Studiorum\\
  Bologna, Italy \\
  \texttt{marco.pasini.98@gmail.com} \\
}

\begin{document}
\maketitle

\begin{abstract}
Traditional voice conversion methods rely on parallel recordings of multiple speakers pronouncing the same sentences. For real-world applications however, parallel data is rarely available. We propose MelGAN-VC, a voice conversion method that relies on non-parallel speech data and is able to convert audio signals of arbitrary length from a source voice to a target voice. We firstly compute spectrograms from waveform data and then perform a domain translation using a Generative Adversarial Network (GAN) architecture. An additional siamese network helps preserving speech information in the translation process, without sacrificing the ability to flexibly model the style of the target speaker. We test our framework with a dataset of clean speech recordings, as well as with a collection of noisy real-world speech examples. Finally, we apply the same method to perform music style transfer, translating arbitrarily long music samples from one genre to another, and showing that our framework is flexible and can be used for audio manipulation applications different from voice conversion.
\end{abstract}


\section{Introduction}
Voice Conversion (VC) is a technique used to change the perceived identity of a source speaker to that of a target speaker, while maintaining linguistic information unchanged. It has many potential applications, such as generating new voices for TTS (Text-To-Speech) systems \cite{Kain1998}, dubbing in movies and videogames, speaking assistance \cite{Kain2007,Nakamura2012} and speech enhancement \cite{Inanoglu2009,Turk2010,Toda2012}. This technique consists in creating a mapping function between voice features of two or more speakers. Many ways of obtaining this result have been explored: Gaussian mixture models (GMM) \cite{Stylianou1998,Toda2007,Helander2010}, restricted Boltzmann machine (RBM) \cite{Chen2014,Nakashika2014}, feed forward neural networks
(NN) \cite{Sun2015}, recurrent neural networks (RNN) \cite{Desai2010,Mohammadi2014} and convolutional neural networks (CNN) \cite{Kaneko2017}. The majority of the mentioned VC approaches make use of parallel speech data, or, in other terms, recordings of different speakers speaking the same utterances. When the recordings are not perfectly time aligned, these techniques require some sort of automatic time alignment of the speech data between the different speakers, which can often be tricky and not robust. Methods that don't require parallel data have also been explored: some of them need a high degree of supervision \cite{Dong2015,MengZhang2008}, requiring transcripts of every speech recording and lacking the ability to accurately capture non-verbal information. Methods involving Generative Adversarial Networks \cite{Goodfellow2014} have also been proposed \cite{Kaneko2018,Kameoka2018,Kaneko2019}: while often producing realistic results, they only allow the conversion of speech samples with a fixed or maximum length. We propose a voice conversion method that doesn't rely on parallel speech recordings and other kinds of supervision and is able to convert samples of arbitrary length. It consists of a Generative Adversarial Network architecture, made of a single generator and discriminator. The generator takes high definition spectrograms of speaker A as input and converts them to spectrograms of speaker B. A siamese network is used to maintain linguistic information during the conversion by the generator. An identity loss is also used to strengthen the linguistic connection between the source and generated samples. We are able to translate spectrograms with a time axis that is arbitrarily long. To accomplish that, we split spectrograms along the time axis, feed the resulting samples to the generator, concatenate pairs of generated samples along the time axis and feed them to the discriminator. This allows us to obtain a generated concatenated spectrogram that doesn't present any discontinuities in the concatenated edges. We finally show that the same technique previously described can also translate a music sample of one genre to another genre, proving that the algorithm is flexible enough to perform different kinds of audio style transfer.

\section{Related Work}
Generative Adversarial Networks (GANs, \cite{Goodfellow2014}) have been especially used in the context of image generation and image-to-image translation \cite{Radford2016,Karras2018,Brock2019,Isola2017,Zhu2017,Liu2017,Liu2019}. Applying the same GAN architectures designed for images to other kinds of data such as audio data is possible and has been explored before. \cite{Donahue2018a} shows that generating audio with GANs is possible, using a convolutional architecture on waveform data and spectrogram data. \cite{Kaneko2018,Kameoka2018,Kaneko2019} propose to use the different GAN architectures to perform voice conversion translating different features (MCEPs, log $F_{0}$, APs) instead of spectrograms. We choose a single generator and discriminator architecture as explained in \cite{Amodio2019}, where a siamese network is also used to preserve content information in the translation. The proposed TraVeL loss aims at making the generator preserve vector arithmetic in the latent space produced by the siamese network. The transformation vector between images (or spectrograms) of the source domain (or speaker) must be the same as the transformation vector between the same images converted by the generator to the target domain. In this way the network doesn't rely on pixel-wise differences (cycle-consistency constraint) and proves to be more flexible at translating between domains with substantially different low-level pixel features. This is particularly effective on the conversion of speech spectrograms, which can be quite visibly different from speaker to speaker. Furthermore, not relying on pixel-wise constraints we are also able to work with audio data different from speech such as music, translating between totally different music genres.

\begin{figure}[t]
\centering
\includegraphics[width=0.85\textwidth]{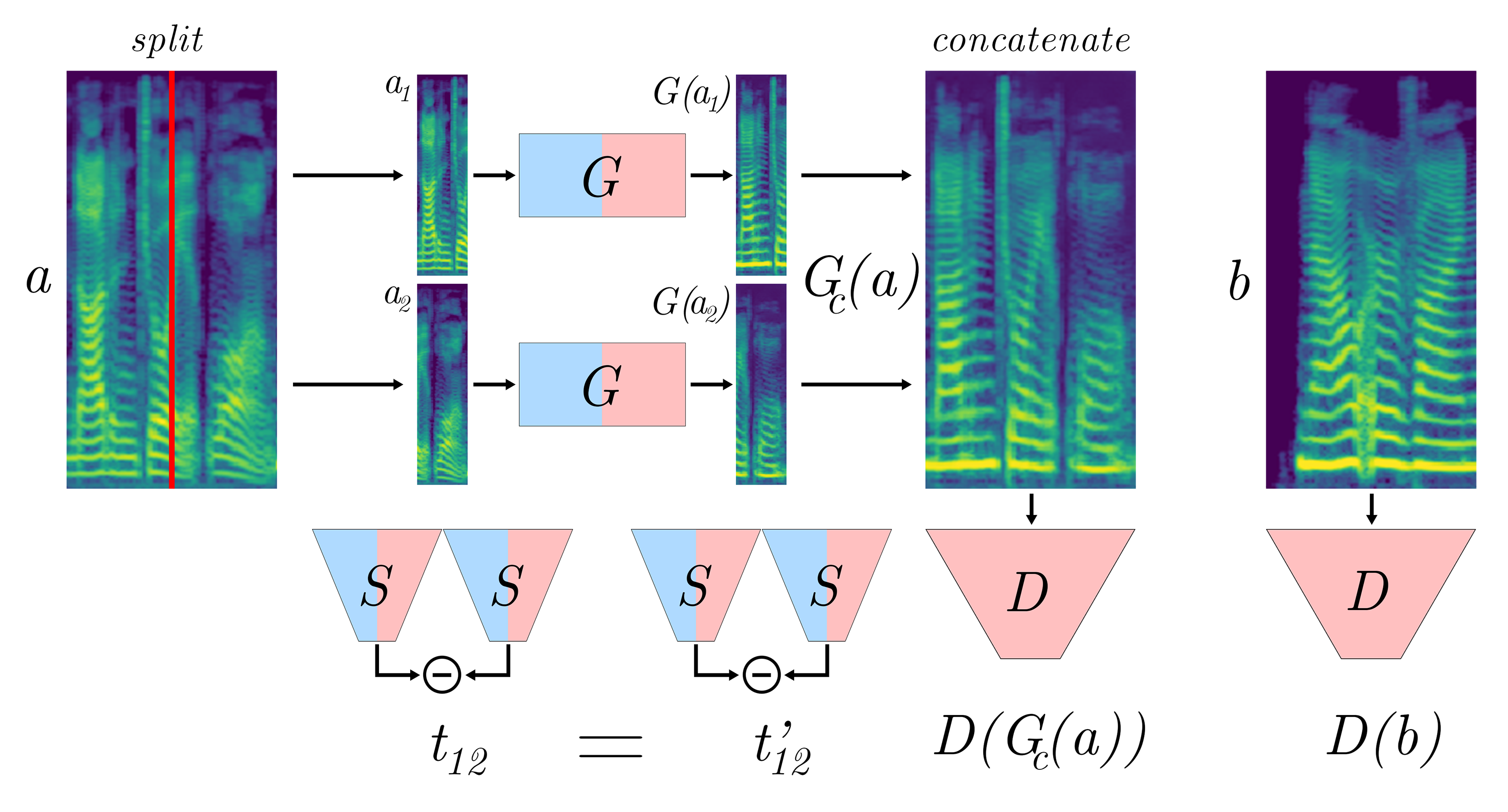}
\caption{MelGAN-VC training procedure. We split spectrogram samples, feed them to the generator $G$, concatenate them back together and feed the resulting samples to the discriminator $D$ to allow translation of samples of arbitrary length without discrepancies. We add a siamese network $S$ to the traditional generator-discriminator GAN architecture to preserve vector arithmetic in latent space and thus have a constraint on low-level content in the translation. An optional identity mapping constraint is added in tasks which also need a preservation of high-level information (linguistic information in the case of voice translation).}
\label{fig:model}
\end{figure}

\section{Model}
Given audio samples in the form of spectrograms from a source domain (speaker, music genre), our goal is to generate realistic audio samples of the target domain while keeping content (linguistic information in the case of voice translation) from the original samples (see Fig. \ref{fig:model}).
\subsection{Spectrogram Splitting and Concatenation}
Let $A$ be the source domain and $B$ the target domain. $\{a_{tot,i}\}_{i=1}^{N_atot}\in A$ and $\{b_{tot,i}\}_{i=1}^{N_btot}\in B$ are the spectrogram representations of the audio samples in the training dataset, each with shape $M\times t$, where $M$ represents the height of the spectrogram (mel channels in the case of mel-spectrograms) and where $t$, the time axis, varies from sample to sample. In order to be able to translate spectrograms with a time axis of arbitrary length, we extract from the training spectrograms $\{a_i\}_{i=1}^{N_a}\in A$ and $\{b_i\}_{i=1}^{N_b}\in B$, each with a shape $M\times L$, where $L<t$ $\forall t$ is a constant. We then split each $a_i$ along the time axis obtaining $a_i^1$ and $a_i^2$ with shape $M\times \frac{L}{2}$. Translating each pair $(a_i^1,a_i^2)$ with a generator $G$ results in pairs $(\hat{b}_i^1,\hat{b}_i^2)$: concatenating them together along the time axis results in $\hat{b}_i$ with shape $M\times L$, where $\{\hat{b_i}\}_{i=1}^{N_a}\in \hat{B}\equiv G(A)$. We finally feed the real samples $\{b_i\}_{i=1}^{N_b}$ and the generated and concatenated samples $\{\hat{b_i}\}_{i=1}^{N_a}$ to a discriminator $D$. With this technique the generator is forced to generate realistic $M\times \frac{L}{2}$ samples with no discontinuities on the edges of the frequency axes, so that when concatenated with adjacent spectrogram samples the final $M\times L$ spectrograms look realistic to the discriminator. After training, when translating a $M\times t$ spectrogram, we first split it in sequential $M\times \frac{L}{2}$ samples (we use padding if $t$ is not a multiple of $\frac{L}{2}$), feed them to the generator and concatenate them back together into the original $M\times t$ shape.
\subsection{Adversarial Loss}
MelGAN-VC relies on a generator $G$ and discriminator $D$. $G$ is a mapping function from the distribution $A$ to the distribution $B$. We call $\hat{B}=G(A)$ the generated distribution. $D$ distinguishes between the real $B$ and the generated $\hat{B}$.  Given training samples $\{a_i\}_{i=1}^{N_a}\in A$ and $\{b_i\}_{i=1}^{N_b}\in B$ with shape $M\times L$ the generator must learn the mapping function. With $G_c(x)$ we define the function that takes the spectrogram $x$ with shape $M\times L$ as input, splits it along the time axis into $(x_{\frac{L}{2}}^1,x_{\frac{L}{2}}^2)$ with shape $M\times \frac{L}{2}$, feeds each one of the two samples to $G$ and concatenates the outputs to obtain a final $M\times L$ spectrogram. An adversarial loss is used: we notice that the hinge loss \cite{Zhang2018} performs well for this task. Thus we use the following adversarial losses for $D$ and $G$
\begin{equation}
\mathcal{L}_{D,adv}=-\mathbb{E}_{b\sim B}[min(0,-1+D(b))]-\mathbb{E}_{a\sim A}[min(0,-1-D(G_c(a)))]
\end{equation}
\begin{equation}
\mathcal{L}_{G,adv}=-\mathbb{E}_{a\sim A}D(G_c(a))
\end{equation}
The discriminator $D$ iteratively learns how to distinguish real samples of distribution $B$ from generated samples of distribution $\hat{B}$, while the generator $G$ iteratively learns how to improve its mapping to increase the loss of $D$. In this way $G$ generates the distribution $\hat{B}$ as similar to $B$ as possible, achieving realism in the generated samples.
\subsection{TraVeL Loss}
Originally introduced in \cite{Amodio2019}, the TraVeL loss (Transformation Vector Learning loss) aims at keeping transformation vectors between encodings of pairs of samples $[(a_{\frac{L}{2},i},a_{\frac{L}{2},j}), (\hat{b}_{\frac{L}{2},i},\hat{b}_{\frac{L}{2},j})]$ equal, $\{G(a_{\frac{L}{2},i})\}_{i=1}^{2N_a}=\{\hat{b_{\frac{L}{2},i}}\}_{i=1}^{2N_a}\in \hat{B}$ being the generated samples with shape $M\times \frac{L}{2}$. This allows the generator to preserve content in the translation without relying on pixel-wise losses such as the cycle-consistency constraint \cite{Zhu2017}, as this makes the translation between complex and heterogeneous domains substantially difficult.We define a transformation vector between $(x_i,x_j)\in X$ as
\begin{equation}
t(x_i,x_j)=x_j - x_i
\end{equation}
We use a cooperative siamese network $S$ to encode samples in a semantic latent space and formulate a loss to preserve vector arithmetic in the space such as
\begin{equation}
t(S(a_{\frac{L}{2},i}),S(a_{\frac{L}{2},j}))=t(S(\hat{b}_{\frac{L}{2},i}),S(\hat{b}_{\frac{L}{2},j}))\quad \forall i,j
\end{equation}
where $S(x)$ with $x\in X$ is the output vector of siamese network $S$.
Thus the loss is the following
\begin{equation}
\mathcal{L}_{(G,S),TraVeL}=\mathbb{E}_{(a_{\frac{L}{2},1},a_{\frac{L}{2},2})\sim A}[cosine\_similarity(t_{12},t'_{12})+|| t_{12}-t'_{12}|| _2^2)] \quad with\ a_{\frac{L}{2},1}\neq a_{\frac{L}{2},2}
\end{equation}
\begin{equation*}
t_{ij}=S(a_{\frac{L}{2},i})-S(a_{\frac{L}{2},j})
\end{equation*}
\begin{equation*}
t'_{ij}=S(G(a_{\frac{L}{2},i}))-S(G(a_{\frac{L}{2},j}))
\end{equation*}
We consider both cosine similarity and euclidean distance so that both angle and magnitude of transformation vectors must be preserved in latent space. The TraVeL loss is minimized by both $G$ and $S$: the two networks must 'cooperate' to satisfy the loss requirement. $S$ can learn a trivial function to satisfy (5), thus we add the standard siamese margin-based contrastive loss \cite{Melekhov2016,Ong2017} to eliminate this possibility.
\begin{equation}
\mathcal{L}_{S,margin}=\mathbb{E}_{(a_{\frac{L}{2},1},a_{\frac{L}{2},2})\sim A}max(0,(\delta-||t_{12}||_2)) \quad with\ a_{\frac{L}{2},1}\neq a_{\frac{L}{2},2}
\end{equation}
where $\delta$ is a fixed value. With this constraint, $S$ encodes samples so that in latent space each encoding must be at least $\delta$ apart from every other encoding, saving the network from collapsing into a trivial function.
\subsection{Identity Mapping}
We notice that when training the system for a voice conversion task with the constraints explained above, while the generated voices sound realistic, some linguistic information is lost during the translation process. This is to be expected given the reconstruction flexibility of the generator under the TraVeL constraint. We extract $b_i^{id}$ samples of shape $M\times \frac{L}{2}$ from original $M\times t$ spectrograms in domain $B$ and we adopt an identity mapping \cite{Taigman2017,Zhu2017} to solve this issue.
\begin{equation}
\mathcal{L}_{G,id}=\mathbb{E}_{b^{id}\sim B}[||G(b^{id})-b^{id}||_2^2]
\end{equation}
The identity mapping constraint isn't necessary when training for audio style transfer tasks that are different from voice conversion, as there is no linguistic information to be preserved in the translation.
\subsection{MelGAN-VC Loss}
The final losses for $D$, $G$ and $S$ are the following
\begin{equation}
\mathcal{L}_D=\mathcal{L}_{D,adv}
\end{equation}
\begin{equation}
\mathcal{L}_G=\mathcal{L}_{G,adv}+\alpha\mathcal{L}_{G,id}+\beta\mathcal{L}_{(G,S),TraVeL}
\end{equation}
\begin{equation}
\mathcal{L}_S=\beta\mathcal{L}_{(G,S),TraVeL}+\gamma\mathcal{L}_{S,margin}
\end{equation}
While $\mathcal{L}_{(G,S),TraVeL}$ aims at making the generator preserve low-level content information, without relying on pixel-wise constraints, $\mathcal{L}_{G,id}$ influences the generator to preserve high-level features. Tweaking the weight constant $\alpha$ allows to balance the two content-preserving constraints. A high value of $\alpha$ will result in generated samples that have similar high-level structure as the source samples, but generally inferior resemblance to the style of the target samples, thus less realistic. On the other hand, eliminating the identity mapping component from the loss ($\alpha=0$) will generally result in more realistic translated samples with less similar structure to the source ones.

\begin{figure}[t]
\centering
\includegraphics[width=0.81\textwidth]{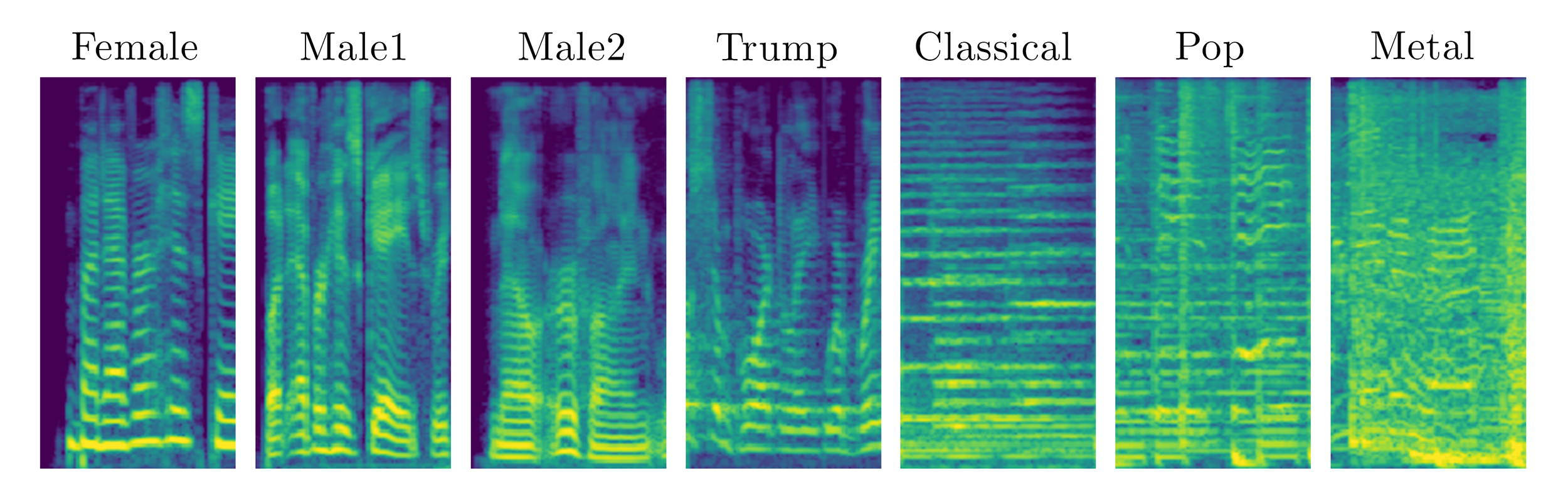}
\caption{Mel-spectrograms (with log-amplitudes) of audio samples. $Female$, $Male1$ and $Male2$ are from the ARCTIC dataset, $Trump$ is from noisy speech samples of Donald Trump extracted from videos on $youtube.com$, while $Classical$, $Pop$ and $Metal$ are from the GTZAN dataset of music samples. Each spectrogram represents $\sim$1 second of audio.}
\label{fig:samples}
\end{figure}

\section{Implementation Details}
We use fully convolutional architectures for the generator ($G$), discriminator ($D$, PatchGAN discriminator \cite{Isola2017}), and siamese ($S$) networks. $S$ outputs a vector of length $len_S$. For our experiments we choose $len_S=128$. $G$ relies on a u-net architecture, with $strides=2$ convolutions for downscaling and sub-pixel convolutions \cite{Shi2016} for upscaling, to eliminate the possibility of checkerboard artifacts of transposed convolutions. Following recent trends in GAN research \cite{Miyato2018,Zhang2018}, each convolutional filter of both $G$ and $D$ is spectrally normalized as this greatly improves training stability. Batch normalization is used in $G$ and $S$. After experimenting with different loss weight values, we choose $\alpha=1$, $\beta=10$, $\gamma=10$ when training for voice conversion tasks, while we eliminate the identity mapping constraint ($\alpha=0$) for any other kind of audio style transfer. During training, we choose $batch\_size=16$, Adam \cite{Kingma2015} as the optimizer, $lr_D=0.0004$ as the learning rate for $D$ and $lr_{G,S}=0.0001$ as the learning rate for $G$ and $S$ \cite{Heusel2017,Zhang2018}, while we update $D$ multiple times for each $G$ and $S$ update. We use audio files with a sampling rate of 16 kHz. We extract spectrograms in the mel scale with log-scaled amplitudes (Fig. \ref{fig:samples}), normalizing them between -1 and 1 to match the output of the $tanh$ activation of the generator. The following hyperparameters are used: $hop\_size=192$, $window\_size=6*hop\_size$, $mel\_channels=hop\_size$, $L=\frac{hop\_size}{2}$. We notice that a higher value of $L$ allows the network to model longer range dependencies, while increasing the computational cost. To invert the mel-spectrograms back into waveform audio the traditional Griffin-Lim algorithm \cite{Griffin1983} is used, which, thanks to the high dimensionality of the spectrograms, doesn't result in a significant loss in quality.

\section{Experiments}
\begin{figure}[t]
\centering
\includegraphics[width=\textwidth]{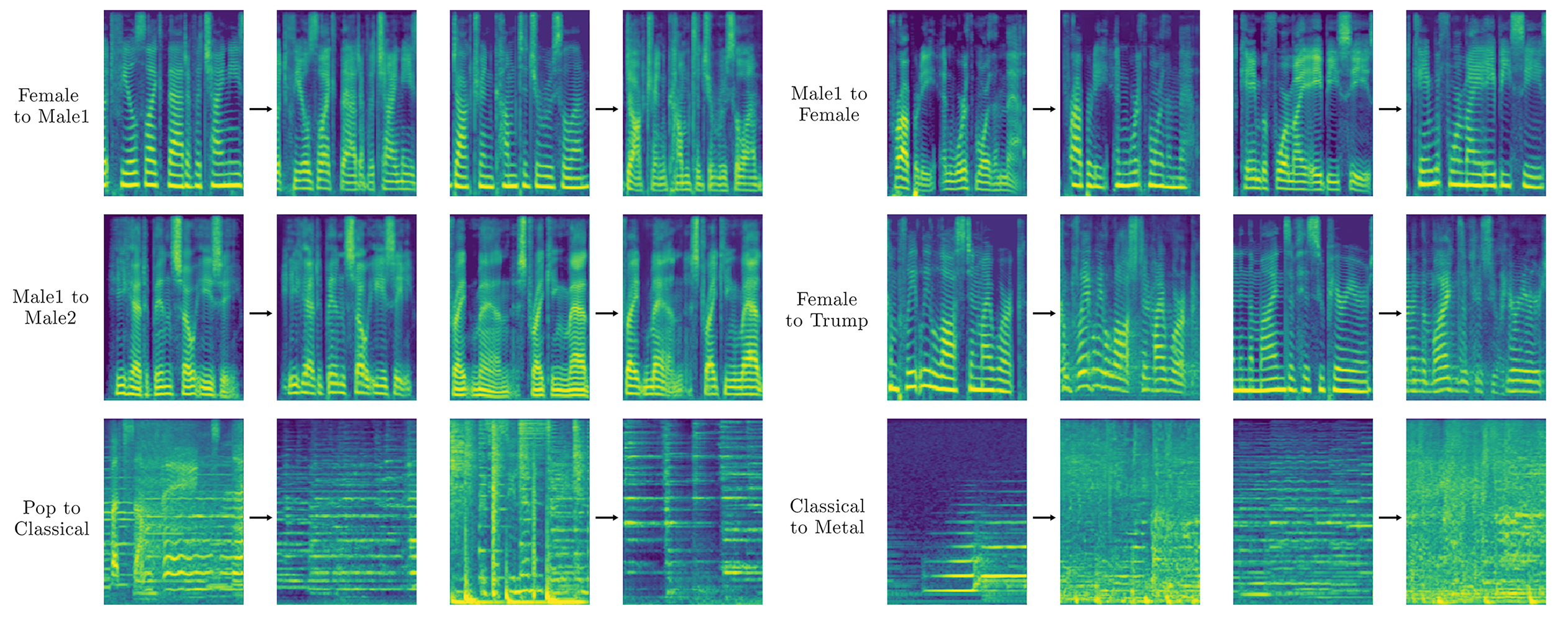}
\caption{Random source (left of each arrow) and translated (right of each arrow) spectrogram samples from multiple categories. Each spectrogram represents $\sim$1.5 seconds of audio. The spectrograms are then converted back to waveform data using the Griffin-Lim algorithm \cite{Griffin1983}.}
\label{fig:converted}
\end{figure}

We experiment with the ARCTIC dataset\footnote{http://www.festvox.org/cmu\_arctic/} for voice conversion tasks. We perform intra-gender and inter-gender voice translation. In both cases MelGAN-VC produces realistic results with clearly understandable linguistic information that is preserved in the translation. We also extract audio from a number of online videos from $youtube.com$ featuring speeches of Donald Trump. The extracted audio data appears noisier and more heterogeneous than the speech data from the ARCTIC dataset, as the Donald Trump speeches were recorded in multiple different real-world conditions. Training MelGAN-VC for voice translation using the real-world noisy data as source or target predictably results in noisier translated speeches with less understandable linguistic information, but the final generated voices are overall realistic. We finally experiment with the GTZAN dataset\footnote{http://marsyas.info/downloads/datasets.html}, which contains 30 seconds samples of different musical pieces belonging to multiple genres. We train MelGAN-VC to perform genre conversion between different musical genres (Fig. \ref{fig:converted}). After training with and without the identity mapping constraint we conclude that it is not necessary for this task, where high-level information is less important, and we decide not to implement it during the rest of our experiments in genre conversion, as this greatly reduces computational costs. If implemented however, we notice that the translated music samples have a stronger resemblance to the source ones, and in some applications this result could be preferred. Translated samples of speech and music are available on $youtube.com$\footnote{https://youtu.be/3BN577LK62Y}. 

\section{Conclusions}
We proposed a method to perform voice translation and other kinds of audio style transfer that doesn't rely on parallel data and is able to translate samples of arbitrary length. The generator-discriminator architecture and the adversarial constraint result in highly realistic generated samples, while the TraVeL loss shows to be an effective constraint to preserve content in the translation, while not relying on cycle-consistency. We conducted experiments and showed the flexibility of our method with substantially different tasks. We believe it is important to discuss the possibility of misuse of our technique, especially given the level of realism achievable by our technique as well as other methods. While applications such as music genre conversion don't appear to present dangerous uses, voice conversion can be easily misused to create fake audio data for political or personal reasons. It is crucial to also invest resources into developing methods to recognize fake audio data.

\bibliographystyle{unsrt}  
\bibliography{references}

\end{document}